\documentclass[conference]{IEEEtran}
\IEEEoverridecommandlockouts
\usepackage{cite}
\usepackage{amsmath,amssymb,amsfonts}
\usepackage{algorithmic}
\usepackage[pdftex]{graphicx}
\usepackage{textcomp}
\usepackage{amsthm}
\usepackage{array}
\usepackage{hyperref}
\usepackage{caption}
\usepackage{soul}
\usepackage{color}
\usepackage{subcaption}
\usepackage{enumitem}
\usepackage{cite}
\usepackage{subfiles}
\usepackage{verbatim}
\usepackage{bbm}
\newtheorem{theorem}{Theorem}

\newtheorem{proposition}{Proposition}
\newtheorem{lemma}{Lemma}
\usepackage[font = small]{caption}
\newcommand{\EE}{\mathbb E}
\newcommand{\cP}{\mathcal P}
\usepackage{xcolor}
\def\BibTeX{{\rm B\kern-.05em{\sc i\kern-.025em b}\kern-.08em
    T\kern-.1667em\lower.7ex\hbox{E}\kern-.125emX}}
\begin{document}

\title{Co-optimizing Consumption and EV Charging under Net Energy Metering}

\author{Minjae~Jeon,~\IEEEmembership{Student Member,~IEEE,}
        Lang~Tong,~\IEEEmembership{Fellow,~IEEE,}
        Qing~Zhao,~\IEEEmembership{Fellow,~IEEE}
        
    \thanks{Minjae Jeon, Lang Tong, and Qing Zhao(\{\textcolor{blue}{\texttt{mj444, lt35, qz16}}\}\textcolor{blue}{\texttt{@cornell.edu}}) are with the School of Electrical and Computer Engineering, Cornell University, USA.}
    }

\maketitle

\begin{abstract}
We consider the co-optimization of flexible household consumption, electric vehicle charging, and behind-the-meter distributed energy resources under the net energy metering tariff. Using a stochastic dynamic programming formulation, we show that the optimal co-optimization follows a procrastination threshold policy that delays and minimizes electricity purchasing for EV charging. The policy thresholds can be computed off-line, simplifying the continuous action space dynamic optimization. Empirical studies using renewable, consumption, and EV data demonstrate 30 \% and 65 \% improvements in customer surplus over the state-of-the-art for the typical 6 and 8-hour scheduling horizons, respectively. The impacts of net energy metering parameters on the household surplus are also demonstrated.
\end{abstract}


\section{Introduction}
We address the problem of co-optimizing electric vehicle (EV) charging and flexible consumption decisions of a prosumer in a household with behind-the-meter (BTM) distributed energy resources (DER).   Because  EV charging represents a substantial part of the total energy consumption and is deferrable, coordinating EV charging with available BTM DER and other flexible demands can reduce the energy cost and increase the overall economic benefits for the prosumer. 

\begin{figure}[t]
    \centerline{\includegraphics[width = \columnwidth]{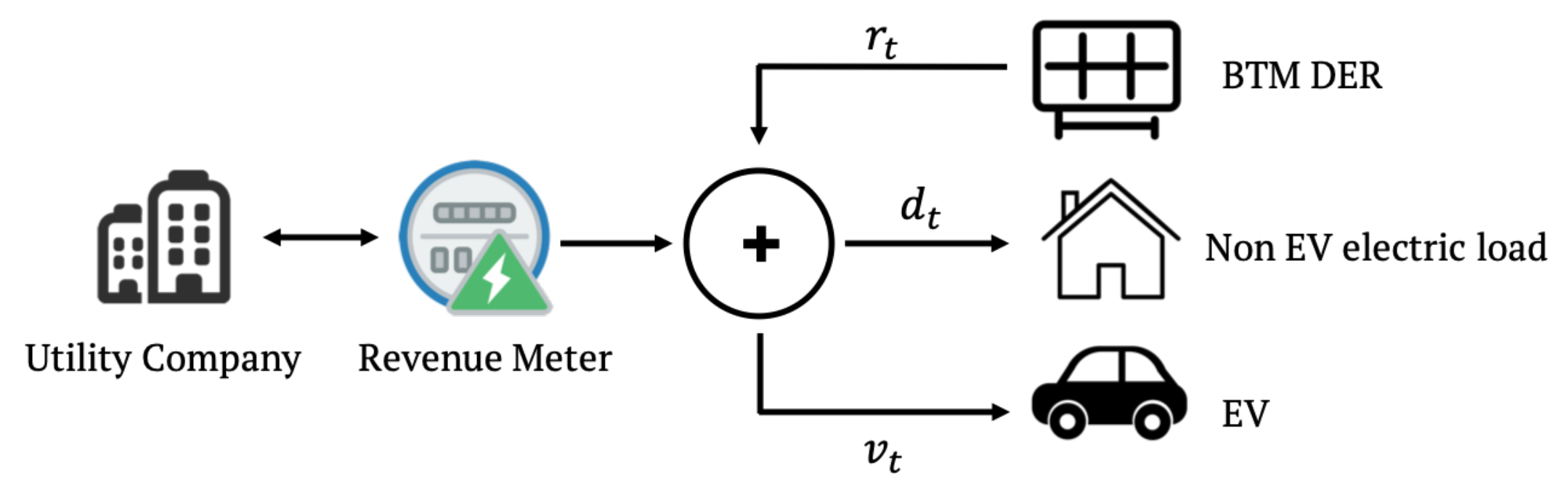}}
    \caption{NEM scheme for the household with EV. Direction of arrow indicates positive direction of energy flow.}
    \label{fig1}
    \vspace{-1em}
    \end{figure}

Fig.\ref{fig1} illustrates a generic smart home where a home energy management system (EMS) schedules EV charging and flexible demands such as HVAC\footnote{Heating, ventilation, and air conditioning} and major appliances based on the available BTM DER generation and the price of electricity.

We assume that the household is under its utility's net energy metering (NEM) tariff, where the customer is billed for its {\em net consumption} based on the reading of the revenue meter. The BTM DER is stochastic, and the EMS has access to the renewable generation up to the time of scheduling decision such that EV charging and consumption can be scheduled based on the available renewables.  The co-optimization objective of the EMS is to schedule EV charging and (flexible) household demand in real-time to satisfy the charging demand of EV and maximize the expected accumulative prosumer surplus over a specific scheduling horizon over a day, a week, or longer.

The co-optimization problem considered here falls in the general category of stochastic dynamic programming (DP) in a continuous action space that defines the EV charging and household consumption decisions.  In general, such an infinite-dimensional optimization is intractable unless the problem has particular structures to exploit.  To this end, we examine the structure of time-of-use NEM pricing that has been shown to offer highly structured optimal decisions for certain utility maximization problems \cite{alahmed2022net}.  Our aim is to find a low-complexity solution that achieves optimality under mild assumptions.

\subsection{Related works}
    The literature on home EMS with EV charging is vast. To our best knowledge, however, no existing literature tackles co-optimizing EV charging, household consumption, and BTM renewables under the NEM tariffs. Here we review closely related works on EV charging and consumption co-optimization and the use of DP for co-optimizing EV charging with DER.
    
    Most relevant to our approach is the co-optimization of EV charging and consumption for home EMS\cite{chen2012real, nguyen2013joint} without involving the NEM tariffs. Both approaches are based on one-shot scheduling models that do not adapt to real-time renewable measurements. The mixed integer linear programming (MILP) used in \cite{chen2012real, nguyen2013joint} also have exponentially growing complexity with the length of the scheduling horizon.  Under the wholesale stochastic prices, the model predictive control (MPC) algorithm was adopted for studying the co-optimization of EV and HVAC with uncertainties \cite{chen2013mpc, yousefi2020predictive}. While the MPC approach can adapt in real-time locally available renewables, its performance is affected by the forecast accuracy of future demand, and it has exponentially growing complexity with the look-ahead forecast horizon. Co-optimization using reinforcement learning was studied in \cite{wu2018optimizing, xu2020multi}. In \cite{xu2020multi}, the authors used a neural network and Q-learning to learn uncertainties with the wholesale electricity prices and BTM solar generation. A significant issue of EV charging models adopted in \cite{chen2012real,nguyen2013joint, chen2013mpc, yousefi2020predictive, wu2018optimizing, xu2020multi} is that they do not capture the deferrable nature of EV charging when there are charging completion deadlines.

	There is substantial literature on using DP for co-optimizing EV charging with stochastic renewables under various settings \cite{wu2016stochastic,hafiz2019coordinated,xu2016dynamic,yu2018deadline, jin2019priority}, although they did not address consumption co-optimization under NEM tariff. Stochastic DP was formulated to co-optimize EV charging with the renewables and demand uncertainties, but consumption was not co-optimized in \cite{wu2016stochastic, hafiz2019coordinated}. Both work didn't explicitly model grid exporting price which is key element in NEM tariff. In \cite{ xu2016dynamic,yu2018deadline, jin2019priority} authors developed simple decision rules to tackle the curse of dimensionality on scheduling multiple EVs. The authors of \cite{xu2016dynamic} proposed priority rules based on the less-laxity longer-remaining-processing-time principle. Whittle's index policy and its asymptotic optimality were derived for the deadline scheduling problem in \cite{yu2018deadline}.

\subsection{Summary of results and contributions}
    \begin{itemize}
        \item We show that the optimal scheduling policy is a {\em procrastination threshold policy} where the optimal EV charging policy in every interval is to delay if possible and charge a minimal amount if necessary. This policy is characterized by the procrastination thresholds that can be computed offline when the BTM DER is modeled as a sequence of independent random variables.
        
        \item  We show that, similar to the consumption scheduling problem without EV charging \cite{alahmed2022net}, the optimal net consumption from the co-optimization is a monotonically decreasing piecewise linear function that has three zones on the axis of the available BTM DER, specified by two thresholds. In particular, when the DER is below or above the two thresholds total consumption including EV charging is a constant, and net consumption decreases linearly. When the DER level is in between two thresholds, total consumption including EV charging is matched with the current DER generation and net consumption is zero. Such a net-zero demand zone reduces the overall (especially reverse) power flow in the distribution grid.
        \item The empirical studies using actual data sets demonstrate the benefits of the procrastination policy over the existing method, achieving 30 and 65 \% increased surplus when the horizon length is 8 and 12 hours, respectively.
    \end{itemize}

The notations used in the paper are standard. Vectors are in boldface, with $(x_1, \ldots, x_N)$ a column vector. We use $\mathbf 1$ for a column of ones with appropriate size. $\mathbbm 1_A$ is an indicator function that maps to 1 if A is true, otherwise to zero. 

The proofs of theorems and propositions are omitted due to space limitations, and they can be found at \cite{jeon2022co_opt}.

\section{Problem Formulation}
We consider a sequential EV charging ($v_t$) and consumption ($d_t$) decision problem over a discrete-time and finite horizon $\mathcal T = \{0, \ldots, T-1\}$. Without loss of generality, we assume that $T$ is the completion deadline for EV charging.  Once EV charging is completed, the co-optimization problem has a simple threshold policy \cite{alahmed2022net}.

\subsection{EV charging and consumption model}

\paragraph{BTM DER $(r_t)$}  BTM DER generation in interval $t$ is an exogenous variable modeled as a sequence of independent random variables with the distribution $f_t(\cdot)$. 
\paragraph{Remaining charging demand $(y_t)$ and constraints}  At the beginning of each interval, the remaining charging demand $(y_t)$ is measured. In a time slot, a charger with maximum capacity $\bar v$ and charging efficiency $\eta$ supplies $v_t$. EV is not allowed to discharge and EV is not charged beyond what’s requested at the beginning of the charging session (i.e.$ \; y_t \ge 0$).
\begin{gather}
    y_{t+1} = y_t - \eta v_t,\quad t \in \mathcal T.\\
    v_t \in [0, \min \{y_t / \eta, \, \bar v\}], \quad t \in \mathcal T.
\end{gather}
Without loss of generality we assume that $\eta = 1$ \footnote{Rescale $y_t$ by $1/\eta$.}
\paragraph{Household Consumption $(\mathbf d_t)$}  We model flexible consumption of $K$ controllable devices in interval $t$ by a consumption vector $\mathbf d_t=\big(d_t^{(1)},\cdots,d_t^{(K)}\big)$ where the consumption of device $i$ is constrained by $0\le d_t^{(i)} \le \bar{d}^{(i)}$.

Utility $U_t(\mathbf d_t)$ of consuming $\mathbf d_t$ in interval $t$ is linearly separable concave function with marginal utility $L_t(\mathbf d_t)$ 
\begin{equation*}
    U_t(\mathbf d_t) = \sum_{i=1}^K U_t^{(i)}(d_t^{(i)}), \; L_t(\mathbf d_t) := \nabla U_t = \big(L_t^{(1)}, \ldots, L_t^{(K)}\big) .
\end{equation*}

Uncontrollable loads could be taken into account by subtracting the renewable generation in the current interval and computing adjusted BTM DER distribution.

\paragraph{Net consumption ($z_t$)}  Household net energy consumption in the interval $t$ is the total consumption including EV charging deducted by BTM DER generation.
\begin{equation*}
    z_t := v_t + \mathbf 1 ^T\mathbf d_t - r_t, \quad t \in \mathcal T.
\end{equation*}
We say household is \textit{net consuming} when $z_t > 0$, and \textit{net producing} when $z_t < 0$.

\subsection{NEM TOU tariff model} 
\paragraph{NEM payment $\big(P^{\pi_t}(z_t)\big)$} A household enrolled in the NEM tariff program is billed or credited by the net energy consumption in each billing period. In practice, the billing period ranges from 15 to 30 minute intervals. We matched the length of decision intervals with the billing period which allows us to index the billing period with $t$.

In interval $t$ given NEM tariff parameters $\pi_t = (\pi_t^0, \pi_t^-, \pi_t^+)$, household payment with net consumption $z_t$ is 
\begin{equation}
    P^{\pi_t}(z_t) := z_t(\mathbbm 1_{z_t \ge 0} \pi_t^+ + \mathbbm 1_{z_t \le 0} \pi_t^-)+ \pi^0_t.
\end{equation}
where $\pi_t^+$ is a \textit{retail rate} for the net consumption, $\pi_t^-$ is a \textit{sell rate} for the net generation to the grid and $\pi_t^0$ is the fixed charge. The NEM tariff model was proposed in \cite{alahmed2022net}.\footnote{Fixed charge doesn't affect the optimal decision, so we assume $\pi_t^0 = 0$.}. 

\begin{figure}[t]
\includegraphics[width = \columnwidth]{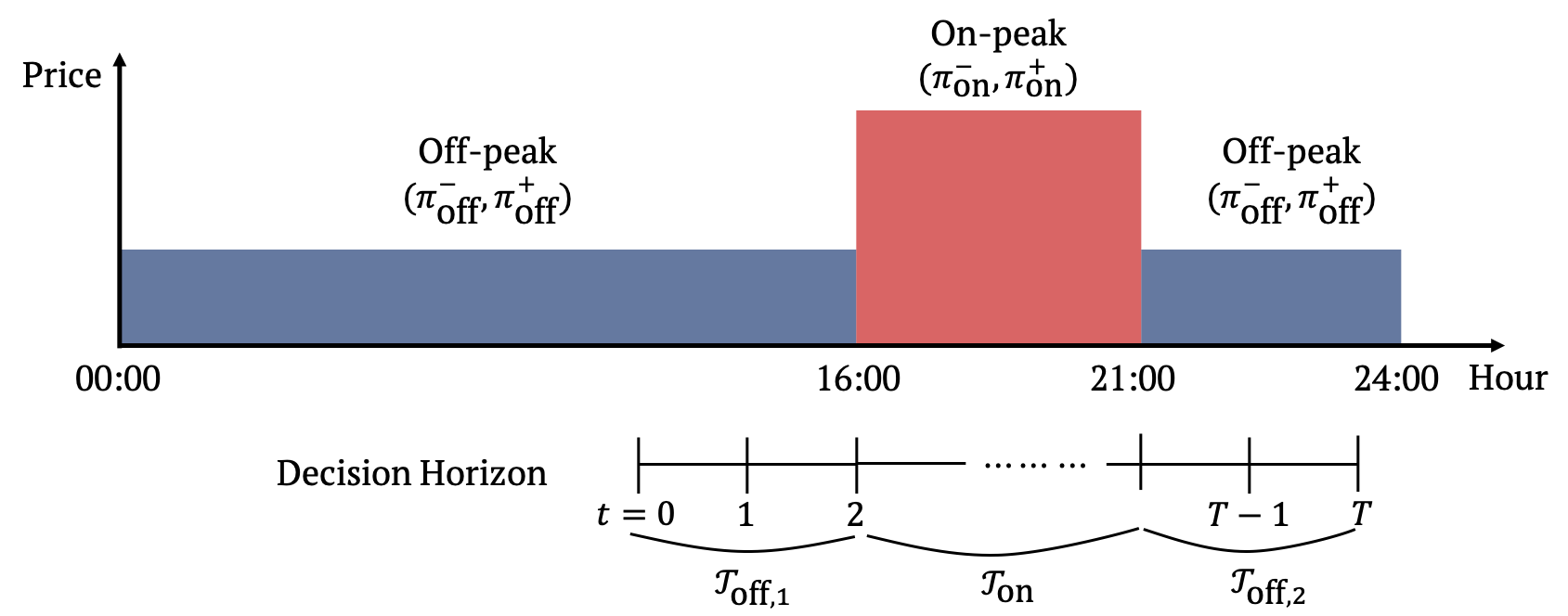}
\caption{TOU scheme and decision horizon}
\label{fig2}
\vspace{-1em}
\end{figure}

\paragraph{NEM TOU tariff model} TOU tariff divides 24 hours into blocks of different prices. We adopt TOU tariff with 2 periods, off-peak and on-peak period, as depicted in Fig.\ref{fig2}. We assume that all the intervals in the same period has the same NEM parameters. Hence, $\pi_t$ is either $\pi_{\text{on}} := [ \pi_{\text{on}}^-, \pi_{\text{on}}^+]$ or $\pi_{\text{off}}:= [ \pi_{\text{off}}^-, \pi_{\text{off}}^+]$. We assume that NEM parameters satisfy $\pi_{\text{off}}^-  <\pi_{\text{on}}^-< \pi_{\text{off}}^+ < \pi_{\text{on}}^+$.

A typical TOU tariff sets the on-peak hours to the six-hour period in the late afternoon and early evening\cite{toureport}. We assume that EV demand falls within the first set of off-periods, the on-periods, and the second set of off-periods as shown in Fig.\ref{fig2}. The case that the EV charging can be completed entirely in the off period becomes a special case. 
Once the starting time for the EV charging and the deadline of completion are known, the scheduling horizon is well defined : $\mathcal T = \mathcal T_{\text{off},1} \cup \mathcal T_{\text{on}} \cup \mathcal T_{\text{off},2}$, with $T = |\mathcal T_{\text{off},1}| + |\mathcal T_{\text{on}}| + |\mathcal T_{\text{off},2}|$. 

\subsection{EV charging consumption co-optimization}
We formulate the co-optimization problem as a stochastic DP. The state of the DP $x_t=(y_t, r_t, \pi_t)$ is defined by the remaining charging demand $y_t$, the available renewable $r_t$, and the NEM tariff parameter $\pi_t$ with $y_0 = s_{\text{req}}$.

Policy $\mu := (\mu_0, \ldots, \mu_{T-1})$ is a sequence of functions that maps state to charging and consumption : $\mu_t(x_t) := (v_t, \mathbf d_t)$.

The stage reward is a household surplus under NEM and terminal reward is a penalty modeled as linear function of incomplete charging demand at the deadline.
\begin{equation*}
    g_t(x_t, v_t, \mathbf d_t) := 
    \begin{cases}U_t(\mathbf d_t) - P^{\pi_t} (z_t), & t = 0, \ldots, T-1. \\
    -\gamma y_T, & t = T.
    \end{cases}
\end{equation*}
Charging and consumption co-optimization is defined by
\begin{equation}
    \begin{aligned}
    \mathcal P : & \max_\mu &&\mathbb E_\mathbf r\left[\sum_{t=0}^{T-1} g_t(x_t, v_t, \mathbf d_t) - \gamma y_T\right] \\
    & \text{s.t.} && \mu_t(x_t) = (v_t, \mathbf d_t) \in \mathcal U(x_t), \quad  \forall t\in\mathcal T  \\
    & && y_{t+1} = y_t - v_t,  \quad \forall t\in\mathcal T  \\
    & && y_t \in [0, s_{\text{req}}],  \quad \forall t\in\mathcal T  \\
    & && y_0 = s_{\text{req}} \\
    & && \pi_t = \pi_{\text{off}}, \quad \forall t \in \mathcal T_{\text{off},1} \cup \mathcal T_{\text{off},2} \\
    & && \pi_t = \pi_{\text{on}},  \quad \forall t \in \mathcal T_{\text{on}} \\
    & && r_t \sim f_t(\cdot), \quad \forall t \in \mathcal T,
    \end{aligned}
\end{equation}
where $\mathcal U(x_t) $ represents the set of feasible actions.

Denote the optimal value function as $V_t(x_t)$ which satisfies the following Bellman Equation : 
\begin{equation}\label{eq:dp}
    V_t(x_t) = \max_{v,\mathbf d} \;\{g_t(x_t, v, \mathbf d) + \mathbb E[V_{t+1}(x_{t+1})] \} 
\end{equation}

\subsection{Model Assumption}
\begin{enumerate}[label = A\arabic*., start = 1]
    \item \textit{High penalty} : Assume that penalty satisfies
    \begin{equation*}
    \pi_{\text{off}}^- < \pi_{\text{on}}^- < \pi_{\text{off}}^+ < \pi_{\text{on}}^+ < \gamma. 
    \end{equation*}
\end{enumerate}
A1 reflects the purpose of minimizing incomplete charging demand.

\section{Optimal EV owner decision}
We now present the optimal EV owner's decision and off-line computation method of the procrastination thresholds.

\subsection{Optimal EV charging policy and consumption policy}
We first present the structural properties of the optimal net consumption, and EV charging and consumption decisions. 
\begin{theorem}[Optimal net consumption] NEM TOU tariff parameters satisfying A1, and $r_t$ is a sequence of independent random variables, optimal net consumption $z_t^*$ is a piecewise linear function of $r_t$ and partitioned into 3 zones, for $t\in \mathcal T$
\begin{gather}
    z_t^* = \begin{cases}
    \Delta_t^+(y_t) - r_t, & r_t \in [0, \Delta_t^+(y_t)\big) \\
    0, & r_t \in [\Delta_t^+(y_t), \Delta_t^-(y_t)] \\
    \Delta_t^-(y_t) - r_t, & r_t \in \big(\Delta_t^-(y_t), \infty),
    \end{cases} \\
    \begin{aligned}
    \text{where }
    \Delta_t^+(y_t) &= \sum_{i=1}^K l_t^{(i)}(\pi_t^+) +v_t^+(y_t), \notag\\
    \Delta_t^-(y_t) &= \sum_{i=1}^K l_t^{(i)}(\pi_t^-)  + v_t^-(y_t), \notag \\
     l_t^{(i)}(\pi) &:= \min\{L_t^{(i)^{-1}}(\pi), \bar  d^{(i)}\}, \notag \\
    v_t^+(y_t) &:= \min \{ \bar v,  \max\{y_t - \tau_t, 0\}\}, \\
    v_t^-(y_t) &:= \min \{ \bar v,  \max\{y_t - \delta_t, 0\}\}.\\
    \end{aligned}
\end{gather}
EV charging and consumption decisions are : \\
If $r_t \in [0,\Delta_t^+(y_t)\big)\,$  for all $i = 1, \ldots, K$
\begin{align}
    v_t^* &= v_t^+(y_t) \label{eq:vplus},\\ 
    d_t^{(i)*} &= l_t^{(i)}(\pi_t^+). \label{eq:dplus}
\end{align}
If $r_t \in \big(\Delta_t^-(y_t), R\big]\,$ for all $i = 1, \ldots, K$
\begin{align}
    v_t^* &= v_t^-(y_t), \label{eq:vminus}\\ 
    d_t^{(i)*} &= l_t^{(i)}(\pi_t^-). \label{eq:dminus}
\end{align}
If $r_t \in [\Delta_t^+(y_t), \Delta_t^-(y_t)]\,$ for all $i = 1, \ldots, K$ 
\begin{flalign}
    \bar V_t(y) &:= \mathbb E[V_t(y, r_t, \pi_t)], \quad h_t(y):= (\partial \bar V_t)(y)  \notag\\
    d_t^{(i)*} &= l_t^{(i)}(\nu) \label{eq:dzero}\\
    v_t^* &= \min \{ \bar v,  \max \{ y_t - h_{t+1}^{-1}(-\nu),0\}\} \label{eq:vzero}\\
    r_t &=v_t^* + \sum _{i=1}^K d_t^{(i)} \label{eq:vdzero},
\end{flalign}
where $\nu \in [\pi_t^-, \pi_t^+]$.
\hfill \qedsymbol
\end{theorem}

\begin{figure}[t]
\centering
\captionsetup[subfigure]{justification=centering}
\begin{subfigure}[t]{.5\columnwidth}
  \centering
  \includegraphics[width = \linewidth]{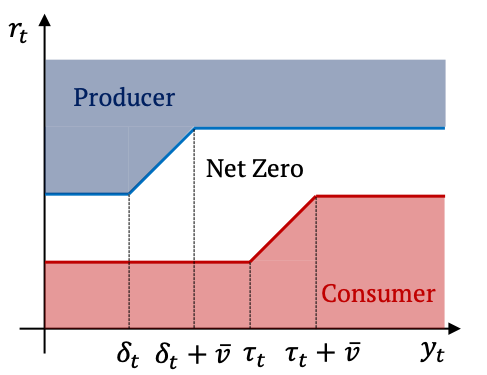}
  \caption{On-peak intervals}
  \label{fig3a}
\end{subfigure}%
\begin{subfigure}[t]{.5\columnwidth}
  \centering
  \includegraphics[width =\linewidth]{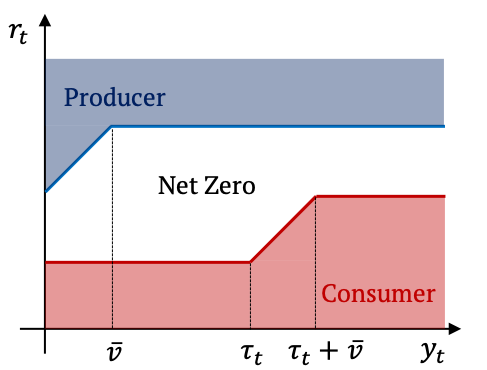}
  \caption{Off-peak intervals}
  \label{fig3b}
\end{subfigure}
\caption{Net consumption regions in the state space}
\label{fig3}
\vspace{-1.2em}
\end{figure}

    State space of $y_t$ and  $r_t$ is divided into 3 zones by $\Delta_t^+(y_t)$ and $\Delta_t^-(y_t)$ as depicted in the Fig.\ref{fig3a}, \ref{fig3b}. Blue and red boundary lines in the figure represent $\Delta_t^-(y_t)$ and $\Delta_t^+(y_t)$, respectively. The theorem suggests that the charging decision is constant for fixed $y_t$ and consumption decisions are constant and charging decisions are constant within the colored region.
    
    The optimal charging decision is the \textit{procrastination threshold policy} which is characterized by the two procrastination thresholds, $\tau_t$ and $\delta_t$, which are the limit of procrastination in which the controller starts charging EV. The procrastination threshold of charging EV using the BTM DER is $\delta_t$ and using energy from the grid is $\tau_t$.
    
    We consider a special case, a charging session that starts and ends within the off-peak period, to gain insight into the procrastination threshold policy. Under the procrastination threshold policy, if it's possible to finish charging requirement in the remaining intervals $\big(y_t \le (T-t-1) \bar v\big)$, the controller delays purchasing for charging and charge only if there is an excess renewable. Otherwise, the controller purchases energy from the grid but a minimal amount that allows completing charging demand in the remaining intervals. Procrastination behavior is optimal in the sense that it increases the probability of using renewables to complete the charging requirement.
    
     The expected marginal value of charging at the interval $t$ with the remaining charging demand $y$ is $h_{t+1}(y)$. Expected marginal value of charging is the expected price of energy we can save by charging at the current interval. At the net-zero zone, charging decision is made so that expected marginal value is matched to the marginal utility of consumption.  

\subsection{Computation of procrastination thresholds}
We now present an off-line computation method and properties of the procrastination threshold. We first present the recursive relation of the thresholds.

\begin{proposition}[Recursive relation of procrastination threshold]
	NEM TOU tariff parameters satisfying A1, and DER modeled as a sequence of independent random variables, within the same period, $\tau_t,\, \tau_{t+1}$ and $\delta_t, \delta_{t+1}$ satisfy
	\begin{flalign*}
	    &&\tau_t &= \tau_{t+1} + \bar v&& \\
	    &&\delta_t &= \delta_{t+1}.&&\qedsymbol
	\end{flalign*}

\end{proposition}

    The proposition implies that it suffices to find procrastination thresholds at the last interval of each pricing period to determine the procrastination thresholds at every interval. 

\begin{theorem}[Optimal procrastination threshold]
NEM TOU tariff parameter satisfying A1, and DER modeled as a sequence of independent random variables, $\tau_t$ and $\delta_t$ are 
\begin{align*}
\tau_t :& \
\begin{array}{cc}
     \tau_t = (T-t-1) \bar v & t \in \mathcal T_{\mathrm{off},2} \cup \mathcal T_{\mathrm{on}}  \\
     \tau_t \text{ s.t. } \pi_{\mathrm{off}}^+ = -h_{t+1}(\tau_t)& t \in \mathcal T_{\mathrm{off},1}
\end{array}\\
\delta_t : &\
    \begin{array}{cc}
         \delta_t  = 0 & t \in \mathcal T_{\mathrm{off},1}\cup \mathcal T_{\mathrm{off},2}   \\
         \delta_t = 0 & t \in \mathcal T_{\mathrm{on}} \land \mathcal T_{\mathrm{off},2} = \emptyset  \\
         \delta_t \text{ s.t. } \pi_{\mathrm{on}}^- = -h_{t+1}(\delta_t) & t \in \mathcal T_{\mathrm{on}} \land \mathcal T_{\mathrm{off},2} \neq \emptyset 
     \end{array}
     \quad \qedsymbol
\end{align*} 
\end{theorem}
    For $\tau_t = (T-t-1) \bar v$ energy is purchased for charging only if it is necessary to complete charging requirement. The theorem also implies that the completion of charging demand is guaranteed whenever it's possible ($s_{\text{req}} \le T \bar v)$.
        
    For $t\in \mathcal T_{\text{off,1}}$, $\tau_t$ depends on the distribution of the future DER generation and $\tau_t$ is the $y_t$ such that the expected price of energy we can by charging at the current interval is $\pi_{\textrm{off}}^+$.
    
    Same arguemnt holds for procrastination threshold in net producing zone.
    
    If we relax assumption A1 and raise the sell rate to the retail rate, consuming the BTM DER is indifferent to purchasing energy from the grid. Hence, the optimal policy will be independent of BTM DER.
	
\section{Numerical results}
    We conducted simulations involving stochastic EV charging and consumption co-optimization with random renewable trajectories, initial charging demands, and connection time.
    
    \subsection{Simulation setting}
    We used the average accumulated household surplus gap between the two policies (in percentage) as the performance measure. Two policies compared are the optimal policy based on Theorem 1-2 and price aware renewable independent open-loop policy. Since our work is the first work to consider co-optimization under the NEM tariff, there doesn't exist a comparable method. Hence we used the renewable independent open-loop policy as a baseline which is the optimal action when the sell rate and the retail rate are matched as in\cite{wu2016stochastic, hafiz2019coordinated}. The baseline policy primarily charges EV in the off-peak intervals and consumption decisions are surplus maximization value under the retail rate.
    
    We conducted 2 different Monte Carlo simulations. 1) Varying the length of the decision horizon from 1 to 14 hours. 2) Varying the price gap between the retail rate and the sell rate from 0.11\$/kWh to 0.25\$/kWh.
    
    The TOU retail rate was from the Pacific Gas and Electronic (PG\&E) TOU tariff data \footnote{PG\&E TOU tariff data can be found at \href{https://www.pge.com/tariffs/electric.shtml}{PGE E TOU-B}}. The on-peak period was 4-9 PM.  
    
    The DER distribution was modeled as a rectified normal distribution. The distribution parameters were estimated using the New York residential solar generation data from the Pecan Street \cite{pecanstreetdata}. We used the charging request data from the Adaptive Charging Network to model the distribution of initial charging demand\cite{lee_acndata_2019}. We assumed that the plug-in time is uniformly distributed.
    
    For simplicity, we considered a single controllable consumption and assumed that the utility is a stationary quadratic function, $U(d) = \alpha d - 1/2 \beta d^2$. The coefficients were estimated using the price and the consumption data \cite{utilityfunc}. The maximum charging capacity of the charger was 3.6kW.
    
     \begin{figure}[t]
         \centering
         \includegraphics[width = 0.8 \columnwidth]{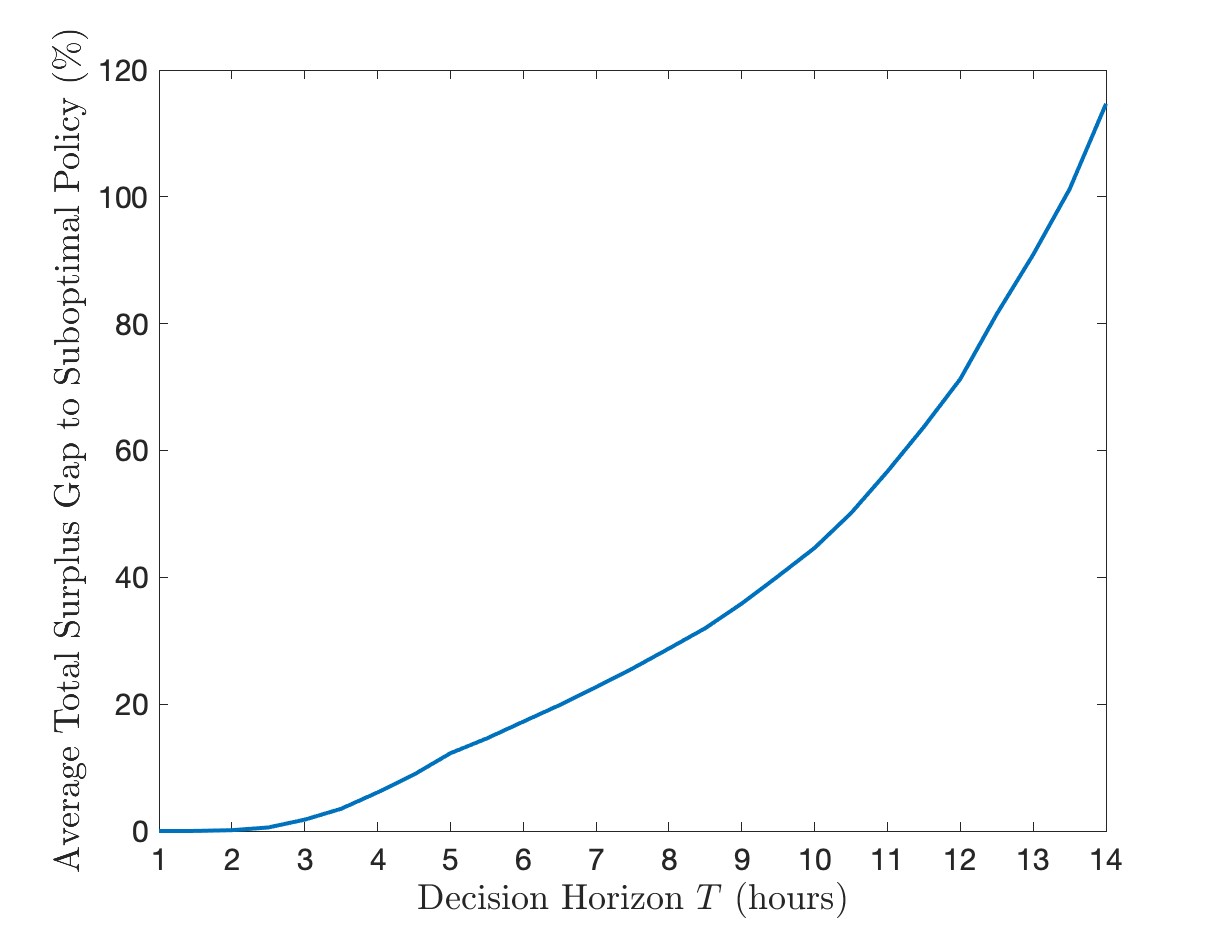}
         \caption{Impact of $T$ on the performance}
         \label{fig7}
         \vspace{-1.5em}
     \end{figure}
     
    \subsection{Influence of $T$}
    Fig.\ref{fig7} shows the influence of the length of the decision horizon on the performance gain. The procrastination threshold policy performs better the longer the charging session is. For the same level of charging demand, the longer charging session provides more flexibility. Hence, it's possible to further procrastinate and complete charging demand using the DER.
    
    \begin{figure}[t]
         \centering
         \includegraphics[width = 0.8 \columnwidth]{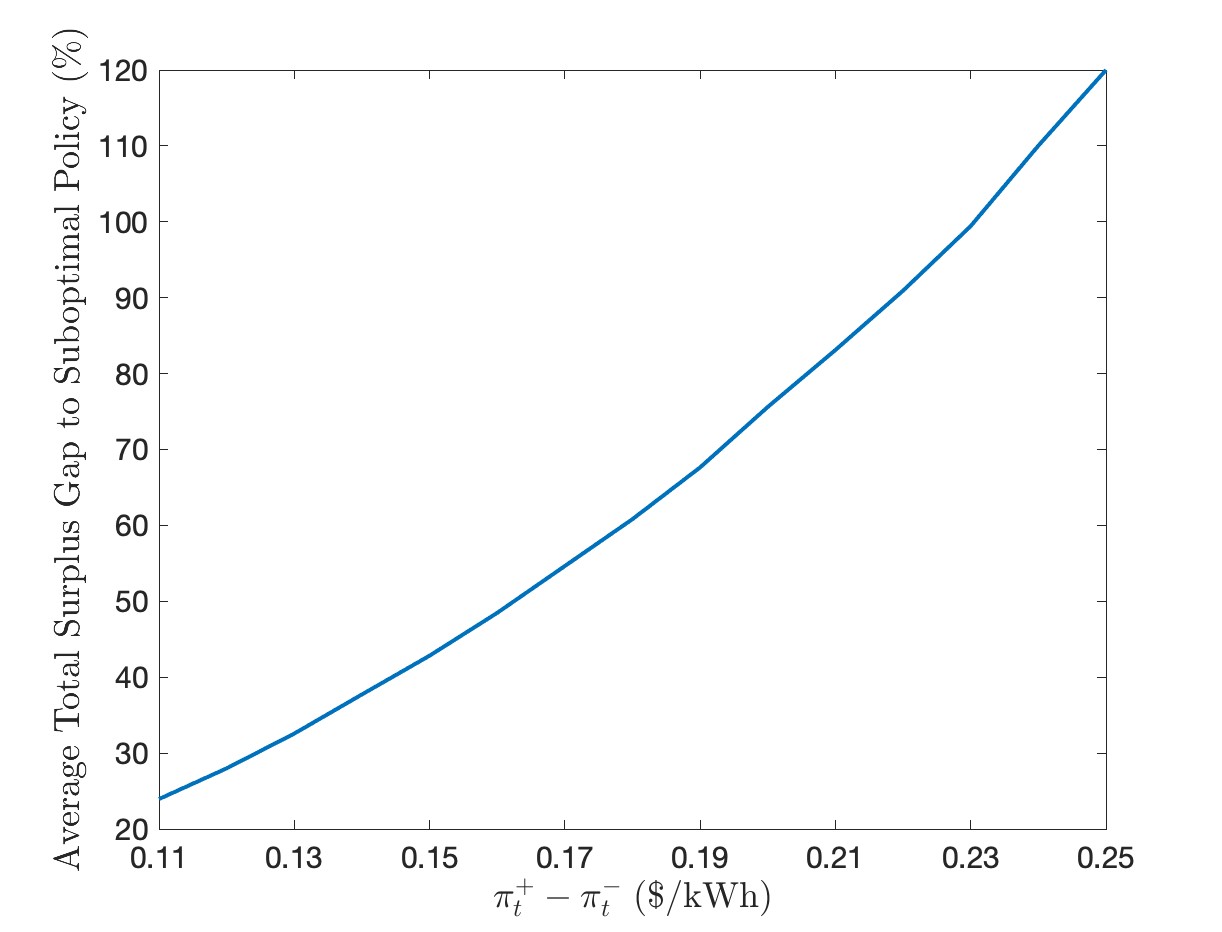}
         \caption{Impact of $\pi_t^+ - \pi_t^- $ on the performance of the optimal policy}
         \label{fig8}
         \vspace{-1.2em}
     \end{figure}
    \subsection{Influence of $\pi_t^+ - \pi_t^-$}
    To observe the effect of the low sell rate on the household surplus, we fixed the retail rate and varied the sell rate assuming that $\pi_t^+ - \pi_t^-$ is preserved for all $t$. The decision horizons are 10 hours. In Fig.\ref{fig8}, the performance gap increases as the price gap increases because lowering $\pi_t^-$ increases utilization of the DER because the value of renewables declines. This result suggests that the procrastination threshold policy has a benefit over the renewable independent policy when the $\pi_t^-$ is further reduced.
    
\section{Conclusion}
EV charging and flexible demand co-optimization is a multi-interval dynamic programming problem that requires optimization in continuous action space.  We show that, under the widely deployed NEM pricing policy, the co-optimization involved has a particular procrastination threshold structure that can be solved easily with predetermined policy thresholds.  

Further analysis on evaluating the impact of different NEM policies can be studied in future works. Our result can be used as a basis to analyze EV owners' responses to different NEM policies. A limitation of our approach is the assumption that the consumption utility and the underlying probability distribution of renewables are known. To this end, inverse reinforcement learning and other machine-learning techniques can be applied in conjunction with the policy structure developed in this work.  

\bibliography{ref.bib}
\bibliographystyle{IEEEtran}
\newpage
\appendices
\section{Proofs}
In the rest of the proofs, for the ease of notations, let's denote $V_t(x_t + \epsilon) = V_t(y_t + \epsilon, r_t, \pi_t)$

Optimal consumption decision without EV charging has been studied in \cite{alahmed2022net}. The theorem that describes two thresholds structure of optimal consumption is shown in \cite{alahmed2022net}.
\begin{theorem}[Prosumer consumption decision under NEM X]
When $\bar v = 0$, the optimal prosumer policy under assumption A1, for every $t \in \mathcal T$ is given by two thresholds $\Delta_t^+, \Delta_t^-$
\begin{align*}
    \Delta_t^+ = \sum_{i=1}^K l_t^{(i)}(\pi_t^+),  \quad \Delta_t^- = \sum_{i=1}^K l_t^{(i)}(\pi_t^-) 
\end{align*}
that divides axis of $r_t$ into 3 zones :
\begin{enumerate}
    \item Net consumption zone : $r_t < \Delta_t^+$. Optimal consumption decision is 
    \begin{equation*}
        d_t^{(i)*} = l_t^{(i)}(\pi_t^+),\; i = 1,\ldots, K
    \end{equation*}
    \item Net producing zone : $r_t > \Delta_t^-$. Optimal consumption decision is 
    \begin{equation*}
        d_t^{(i)*} = l_t^{(i)}(\pi_t^-), \; i = 1, \ldots, K
    \end{equation*}
    \item Net zero zone : $\Delta_t^+ \le r_t \le \Delta_t^-$. Optimal consumption decision is for $\nu \in [\pi_t^-, \pi_t^+]$
    \begin{equation*}
        d_t^{(i)*} = l_t^{(i)}(\nu), \; i = 1, \ldots, K
    \end{equation*}
\end{enumerate}

\end{theorem}

\subsection{Propositions to prove theorems}
\begin{proposition}
Under NEM TOU parameters satisfying A1, $V_t(x_t)$ is monotone decreasing function of $y_t$ and for $\epsilon > 0$, 
\begin{align*}
    -\gamma \epsilon \le V_t(x_t + \epsilon) - V_t(x_t) \le -\pi_{\mathrm{off}}^-\epsilon,
\end{align*}
and for $y_t \ge (T-t) \bar v$, 
\begin{align*}
    V_t(x_t + \epsilon) = V_t(x_t) - \gamma \epsilon,
\end{align*}
for all $t \in \mathcal T$.
\end{proposition}

\begin{proof}
We will prove the proposition using the induction. \\
At $t = T-1$, Bellman equation is
\begin{align*}
    V_{T-1}(x_{T-1}) = \max_{v,\mathbf d} \{g_{T-1}(x_{T-1}, v, \mathbf d) - \gamma (y_{T-1} - v)\} .
\end{align*}
By A1, $V_{T-1}(x_{T-1} + \epsilon)$ satisfies 
\begin{align*}
    V_{T-1}(x_{T-1} + \epsilon) &\le V_{T-1}(x_{T-1}) - \pi_{T-1}^- \epsilon
    \\
    &\le V_{T-1}(x_{T-1}) - \pi_{\mathrm{off}}^-\epsilon 
    < V_{T-1}(x_{T-1}), \\
    V_{T-1}(x_{T-1}) - \gamma \epsilon &\le V_{T-1}(x_{T-1} + \epsilon).
\end{align*}
For $y_{T-1} \ge \bar v$, as $\gamma > \pi_{T-1}^+ > \pi_{T-1}^-$, optimal solution of the Bellman equation is $v^* = \bar v$ and $V_{T-1}(x_{T-1})$ satisfies
\begin{align*}
    V_{T-1}(x_{T-1} + \epsilon) = V_{T-1}(x_{T-1}) - \gamma \epsilon.
\end{align*}
Assuming proposition holds for $t+1$, using linearity and monotonicity of expectation, 
\begin{align*}
    V_t(x_t) - \gamma \epsilon \le V_t(x_t + \epsilon) \le V_t(x_t) - \pi_{\mathrm{off}}^-\epsilon.
\end{align*}
For $y_t \ge (T-t) \bar v$, $y_t - v + \epsilon \ge (T-t-1) \bar v$ for all $v \in [0, \bar v]$,
\begin{equation*}
    V_t(x_t + \epsilon) = V_t(x_t) - \gamma \epsilon.
\end{equation*}
\end{proof}

\begin{proposition}
Under NEM TOU parameters satisfying A1, for all $t \in \mathcal T$, $\EE[V_{t}(x_t)]$ is a concave function of $y_t$.
\end{proposition}
\begin{proof}
Let's prove the proposition using backward induction. At $t = T-1$, there is no randomness with DER generation and the DP equation becomes 
    \begin{equation*}
        \begin{aligned}
            V_{T-1}^*(x_{T-1}) &= \max_{v, \mathbf d} \, \{ g_{T-1}(x_{T-1}, v, \mathbf d) - \gamma (y_{T-1} - v)  \\
            &\text{s.t.} \; v \in [0, \min\{y_{T_1}, \bar v\}] \\
            & \quad \;\, \mathbf d \in [0, \bar {\mathbf d}]
        \end{aligned}
    \end{equation*}

By A1, the optimal EV charging decision is 
\begin{equation*}
    v^* = \min \{ \bar v, y_{T-1}\}.
\end{equation*}
From Proposition 2, for $y_{T-1} \ge \bar v$, $\partial V_{T-1}^*(x_{T_1}) / \partial y_{T-1} = -\gamma$. For $y_{T-1} < \bar v$, optimal consumption decision is decided by EV charging adjusted renewable $\tilde r_{T-1} := \max \{ r_{T-1} - v^*, 0\}$ according to the Theorem 3. As $y_{T-1}$ reduces from $\bar v$, $\tilde r_{T-1}$ increases. By Theorem 3, optimal consumption increases which means marginal utility of consumption increases. Hence, as $y_{T_1}$ increases from 0 to $\bar v$, $\partial V_{T-1}^* (x_{T-1}) / \partial y_{T-1}$ will decrease from $-\pi_{T-1}^-$ to $- \pi_{T-1}^+$. Hence, $\partial V_{T-1}^* (x_{T-1}) / \partial y_{T-1}$ is a decreasing function of $y_{T-1}$, and $V_{T_1}^*(x_{T-1})$ is a concave function of $y_{T-1}$,

Now, suppose $V_{t+1}^*$ is a concave function of $y_{t+1}$. Then, the objective function of the Bellman equation at $t$ is a concave and continuous function of $v$ and $\mathbf d$ in the compact feasible set, there exists a optimal solution for every $y_t$. Let optimal solution for $y_t = y_{t,1}$ as $v_{t,1}^*$ and $\mathbf d_{t,1}^*$, and for $y_t = y_{t,2}$ as $v_{t,2}^*$ and $\mathbf d_{t,2}^*$. For $\lambda \in [0, 1]$, 
    \begin{align*}
        \lambda V_t^*(x_{t,1}) &+ (1 - \lambda) V_t^*(x_{t,2}) \\
        &= \lambda \{g_t(x_{t,1}, v_{t,1}^*, ,\mathbf d_{t,1}^*) + \mathbb E[V_{t+1}^*(x_{t,1} - v_{t,1}^*)]\} \\
        &+(1 - \lambda)\{g_t(x_{t,2}, v_{t,2}^*, ,\mathbf d_{t,2}^*) + \mathbb E[V_{t+1}^*(x_{t,2} - v_{t,2}^*)]\}  \\
        &\le g_t(\xi, v_\xi, e_\xi, \mathbf d_\xi) + \mathbb E[V_{t+1}^*(\xi - v_\xi)] \\
        &\le g_t(\xi, v_\xi^*, e_\xi^*, \mathbf d_\xi^*) + \mathbb E[V_{t+1}^*(\xi - v_\xi^*)] \\
        &= V_t^*(\xi) = V_t^*(\lambda \kappa + (1 - \lambda) \zeta)
    \end{align*}
    Here, $\xi = \lambda x_{t,1} + (1 - \lambda) x_{t,2}$, $v_\xi = \lambda v_{t,1}^* + (1 - \lambda)  v_{t,2}^*$, $\mathbf d_\xi = \lambda \mathbf d_{t,1}^* + (1 - \lambda)  \mathbf d_{t,2}^*$. \\

    The first inequality holds because of the concavity of the stage reward function and inductive hypothesis. The second inequality holds because of the optimality condition. Hence, $V_t^*$ is a concave function of $y_t $ for $t \in \mathcal T$.
\clearpage

\subsection{Proof of Theorem 1}
\begin{proof}
We will divide the Bellman equation (\ref{eq:dp}) into 3 convex optimization problems :  $\mathcal P^+, \mathcal P^-, \mathcal P^0$. 
\begin{align*}
\mathcal P^+ : \; &\max_{v,\mathbf d} && \{ U_t(\mathbf d) - \pi_t^+(v + \mathbf 1^T \mathbf d - r_t) + \EE[V_{t+1}(x_t - v)]\} \\
    & \text{ s.t. }&& v + \mathbf 1 ^T \mathbf d \ge r_t \\ 
    && & \mathbf d \in \mathcal D \\
    & && v \in [0, \min \{\bar v, y_t\}] \\
    \mathcal P^- : \; &\max_{v,\mathbf d} && \{ U_t(\mathbf d) - \pi_t^-(v + \mathbf 1^T \mathbf d - r_t) + \EE[V_{t+1}(x_t - v)]\} \\
    & \text{ s.t. }&& v + \mathbf 1 ^T \mathbf d \le r_t \\ 
    && & \mathbf d \in \mathcal D \\
    & && v \in [0, \min \{\bar v, y_t\}] \\
    \mathcal P^0 : \; &\max_{v,\mathbf d} && \{ U_t(\mathbf d) + \EE[V_{t+1}(x_t - v)]\} \\
    & \text{ s.t. }&& v + \mathbf 1 ^T \mathbf d = r_t \\ 
    && & \mathbf d \in \mathcal D \\
    & && v \in [0, \min \{\bar v, y_t\}] 
\end{align*}
Given $x_t$, $\mathcal P^+, \mathcal P^-, \mathcal P^0$ satisfy Slater's condition. Hence, KKT conditions are the necessary and sufficient condition for the optimality. It suffices to show that (\ref{eq:vplus})-(\ref{eq:vdzero}) satisfy corresponding KKT conditions of $\mathcal P^+, \mathcal P^-, \mathcal P^0$.

Lagrangian of $\mathcal P^+, \cP^-$, and $\cP^0$ are 
    \begin{align*}
        \mathcal L^+ &= U_t(\mathbf d)+(\nu^+ - \pi_t^+)(v + \mathbf 1^T \mathbf d - r_t) + \EE[V_{t+1}(x_t - v)] \\
        & +\underbar{$\lambda$}_v^+ v+ \bar \lambda _v^+(\min\{\bar v, y_t\} - v) + \underbar{$\boldsymbol{\lambda}$}_d^{+T} \mathbf{d} + \bar 
        {\boldsymbol{\lambda}}_d^{+T}(\bar{ \mathbf{d}} - \mathbf{d}),
        \\
        \mathcal L^- &= U_t(\mathbf d)+(\nu^- - \pi_t^+)(v + \mathbf 1^T \mathbf d - r_t) + \EE[V_{t+1}(x_t - v)] \\
        & +\underbar{$\lambda$}_v^- v+ \bar \lambda _v^-(\min\{\bar v, y_t\} - v) + \underbar{$\boldsymbol{\lambda}$}_d^{-T} \mathbf{d} + \bar 
        {\boldsymbol{\lambda}}_d^{-T}(\bar{ \mathbf{d}} - \mathbf{d}),\\
        \mathcal L^0 &= U_t(\mathbf d)+\nu^0 (v + \mathbf 1^T \mathbf d - r_t) + \EE[V_{t+1}(x_t - v)] \\
        & +\underbar{$\lambda$}_v^0 v+ \bar \lambda _v^0(\min\{\bar v, y_t\} - v) + \underbar{$\boldsymbol{\lambda}$}_d^{0T} \mathbf{d} + \bar 
        {\boldsymbol{\lambda}}_d^{0T}(\bar{ \mathbf{d}} - \mathbf{d}),
    \end{align*}
     where $\nu, \underbar{$\boldsymbol{\lambda}$}_d, \bar 
    {\boldsymbol{\lambda}}_d, \bar \lambda _v, \underbar{$\lambda$}_v \ge 0$ are dual variables of the $\cP^+, \cP^-, \cP^0$.

    Let's first verify the optimality of (\ref{eq:dplus}) and (\ref{eq:dminus}). \\
     For $r_t < \Delta_t^+(y_t) $ and $r_t > \Delta_t^-(y_t)$, $\nu^+ = \nu^- = 0$.
    \begin{align}
        \pi_t^+ - \underbar{$\lambda$}_d^{+(i)} + \bar{\lambda }_d^{+(i)} = L_t^{(i)}(d^{(i)})\label{eq:Pplusdplus}. \\
        \pi_t^- - \underbar{$\lambda$}_d^{-(i)} + \bar{\lambda }_d^{-(i)} = L_t^{(i)}(d^{(i)})\label{eq:Pmindmin} .
    \end{align}
    Then, (\ref{eq:dplus}) and (\ref{eq:dminus}) satisfy (\ref{eq:Pplusdplus}) and (\ref{eq:Pmindmin}) for all $i = 1, \ldots, K$.

    Now, let's verify the optimality of (\ref{eq:vplus}) and (\ref{eq:vminus}). By the concavity of $V_t^*$ with respect to $y_t$, $h_t(y)$ is nonempty and monotone. By the Proposition 2, $h_t(y) = -\gamma $ for $y \ge (T-t) \bar v$. Hence, there exist $\tau_t$ and $\delta _t$ in $[0, (T-t-1) \bar v]$ such that $-\pi_t^+ \in h_{t+1}(\tau_t)$ and $-\pi_t^- \in h_{t+1}(\delta_t)$, respectively. Then, (\ref{eq:vplus}) and (\ref{eq:vminus}) satisfy the KKT condition. 

    Similarly, the optimality of (\ref{eq:dzero})-(\ref{eq:vdzero}) can be verified by checking the KKT conditions of $\mathcal P^0$. By the monotonicity of the subdifferential, for $\nu \in [\pi_t^-, \pi_t^+]$, there exists some $\tilde y \in [0, (T-t-1)\bar v]$ that satisfies $-\nu \in h_{t+1}(\tilde y)$. Then, (\ref{eq:dzero})-(\ref{eq:vdzero}) satisfy the KKT condition of $\mathcal P^0$. 

    Also, by the monotonicity of $L_t(\mathbf d)$ and $h_t(y)$, optimal decision in the net zero zone is monotone increasing with respect to $r_t$.
\end{proof}

\begin{lemma}
    Under NEM TOU parameters satisfying A1, for $y_t \ge (T-t) \bar v$, $v_t^* = \bar v$ for all $t \in \mathcal T$ and $d_t^*$ are defined by $\mathrm{Theorem \;3}$ with adjusted renewable $\tilde r_t = \max\{r_t - \bar v , 0\}$.
\end{lemma}
\begin{proof}
We will prove Lemma by contradiction. Suppose $v_t^* < \bar v$. Then, value function is 
\begin{equation*}
    V_t(x_t) = U(\mathbf d_t^*) - P^{\pi_t}(z_t^*) + \EE[V_{t+1}(x_t - v_t^*)]
\end{equation*}
For $\epsilon > 0 $ such that $v_t^* + \epsilon < \bar v$, let value function $\tilde V_t(x_t)$ with same $\mathbf{d}_t^*$ and $v_t^* $ replaced with $v_t^* + \epsilon$. 
\begin{align*}
    \tilde V_t(x_t) &= U(\mathbf d_t^*) - P^{\pi_t}(z_t^* + \epsilon) 
    +\EE[V_{t+1}(x_t - v_t^* - \epsilon)] \\
    & = U(\mathbf d_t^*) - P^{\pi_t}(z_t^* + \epsilon) 
    +\EE[V_{t+1}(x_t - v_t^* )] + \gamma\epsilon \\
    & \ge  U(\mathbf d_t^*) - P^{\pi_t}(z_t^* ) 
    +\EE[V_{t+1}(x_t - v_t^* )] + (\gamma - \pi_t^+)\epsilon\\
    &> V_t(x_t),
\end{align*}
which is a contradiction. Therefore, $v_t^* = \bar v$. As $v_t^* = \bar v$ is independent to the consumption, optimal consumption is defined by BTM DER what's left after charging $\bar v$. The result of optimal consumption decision without EV charging is from Theorem 3.
\end{proof}
In the rest of the proof, denote $\mathbf d^+, \mathbf d^-, \mathbf d^0$ and $v^+, v^-, v^0$ as the optimal solution of $\cP^+, \cP^-, \cP^0,$ respectively.

\subsection{Proof of Theorem 2}

\begin{proof}
From the proof of Theorem 1, $\tau_t$ and $\delta_t$ satisfy $-\pi_t^+ \in h_{t+1}(\tau_t)$ and $-\pi_t^- \in h_{t+1}(\delta_t)$, respectively. Hence, the characterizations of $\tau_t$ for $t \in \mathcal T_{\text{off,1}}$ and $\delta_t$ for $t \in \mathcal T_{\text{on}} \land \mathcal T_{\text{off,2}}$ are proved. 

Now, let's show the special case of procrastination threshold. 

First, for $t \in \mathcal T_{\text{on}} \cup \mathcal T_{\text{off,2}}$ holds, $\pi_{t+1}^+ \le \pi_t^+$, i.e. future purchasing price is non-increasing.

Here, we need to show that $\pi_t^+ \in h_{t+1} \big((T-t-1) \bar v\big)$ for $t \in \mathcal T_{\text{on}} \cup \mathcal T_{\text{off,2}}$. Combining Proposition 2 and property of non increasing retail rate, it's suffice to show that for such $t$ and $\epsilon > 0$,
\begin{equation}\label{eq:convex condition}
    \bar V_{t}\big((T-t)\bar v - \epsilon\big) \le \bar V_{t}\big( (T-t) \bar v \big) +\epsilon \pi_t^+.
\end{equation}

For $t = T-1$, applying Theorem 1 result, we can obtain, 
\begin{align*}
    &\bar V_{T-1}(y_{T-1}) = \Pr( r_{T-1} < \Delta_{T-1}^+) \{U(\mathbf d_{T-1}^+) \\
    &- \pi_{T-1}^+(y_{T-1} + d_{T-1}^+ -\mathbb E [r_{T-1}])\} \\
    &+\Pr(r_{T-1} > \Delta_{T-1}^-) \{ U(\mathbf d_{T-1}^-) -\pi_{T-1}^-(y_{T-1} + d_{T-1}^- \\
    &- \mathbb E[r_{T-1}]\} + \Pr(r_{T-1} \in [\Delta_t^+, \Delta_t^+]) \{U(\mathbf d_{T-1}^0)\}
\end{align*}
Then, $\bar V_{T-1}(\bar v - \epsilon) \le \bar V_{T-1}(\bar v) + \epsilon \pi_{T-1}^+$ because coefficient of $\epsilon$ is a convex combination of terms less than $\pi_{T-1}^+$.

Now, suppose (\ref{eq:convex condition}) holds for $t +1 \in \mathcal T_{\text{on}} \cup \mathcal T_{\text{off,2}}$. Then, by Theorem 1, $\tau_t = (T-t-1) \bar v$ and $v_t^+ = \min \{ \bar v, \max \{ y_t - (T-t-1) \bar v, 0\}\}$ where $v_t^+ \le v_t^0 \le v_t^-$ for fixed $y_t$. 

Applying Theorem 1 to the Bellman equation at $t$ gives
\begin{align*}
    \bar V_t\big((T-t) \bar v - \epsilon\big) \le \bar V_{t+1}\big((T-t-1)\bar v) + \epsilon\pi_t^+ \\
    - \epsilon' \pi_t^+ + \mathbb E[g(x_t, \bar v, \mathbf d^*)] \le  \bar V_t\big((T-t) \bar v\big),
\end{align*}
where $\epsilon' < \epsilon$. The first inequality holds from the monotonicity of $v^*$ with respect to the $r_t$. 
Therefore, (\ref{eq:convex condition}) holds for all $t \in \mathcal T_{\text{on}} \cup \mathcal T_{\text{off,2}}$ and $\tau_t = (T-t-1) \bar v$.

For $t \in \mathcal T_{\text{on}} \land \mathcal T_{\text{off,2}} = \emptyset$. We can prove $\delta_t = 0$ by showing $-\pi_{\text{on}}^- \in h_{t}(0)$. We can prove using the similar argument as we proved characterization of $\tau_t$.

\end{proof}
\subsection{Proof of Proposition 1}
\begin{proof}
For $t \in \mathcal T_{\text{off,2}} \cup \mathcal T_{\text{on}}$, $\tau_t = (T-t-1) \bar v$ satisfies the recursive relation. Hence, it's only necessary to show the recursive relation for $t \in \mathcal T_{\text{off,1}}$. 

For $y_t \ge \tau_t + \bar v$, $v^- = v^+ = \bar v$. By monotonicity $v^0 = \bar v$.  
\begin{equation*}
 h_t(y_t) =  h_{t+1}(y_t - \bar v)   .
\end{equation*}
As $y_t \downarrow \tau_t + \bar v$, $-\pi_{\text{off}}^+ \in h_t(y_t) \to h_{t+1}(\tau_t)$. Since, $h_t(y_t)$ is monotone, $-\pi_{\text{off}}^+ \in h_t(\tau_t + \bar v)  $. Hence, $\tau_{t-1} = \tau_t + \bar v$.

For $t \in \mathcal T_{\text{off,1}} \cup \mathcal T_{\text{off,2}}$ and $t \in \mathcal T_{\text{on}} \land \mathcal T_{\text{off,2}} = \emptyset$, $\delta_t = 0$. Hence, it's only necessary to prove for $ t \in\mathcal T_{\text{on}} \land \mathcal T_{\text{off,2}} \neq \emptyset$. 

For $y_t \le \delta_t$, $v^+ = v^- = 0$. Then, $v^0 = 0$. 
\begin{equation*}
    h_t(y_t) =  h_{t+1}(y_t )   .
\end{equation*}
As $y_t \uparrow \delta_t$, $h_t(y_t) \to h_{t+1}(\delta_t)$. Since $h_t(y_t)$ is monotone, $h_t(\delta_t) = -\pi_{\text{on}}^-$. Hence, $\delta_{t-1} = \delta_t$.
\end{proof}

\end{document}